# *The Group Theoretic Roots of Information: permutations, symmetry, and entropy*

David J. Galas

Pacific Northwest Research Institute

Seattle, Washington, USA

Communication: dgalas@pnri.org

**Abstract**

We propose a new interpretation of measures of information and disorder by connecting these concepts to group theory in a new way. Entropy and group theory are connected here by their common relation to sets of permutations. A combinatorial measure of information and disorder is proposed, in terms of integers and discrete functions, that we call the *integer entropy*. The Shannon measure of information is the limiting case of a richer, more general conceptual structure that reveals relations among finite groups, information, and symmetries. It is shown that the integer entropy converges uniformly to the Shannon entropy when the group includes all permutations, the Symmetric group, and the number of objects increases without bound. The harmonic numbers, $H_n$, have a well-known combinatorial meaning as the expected number of disjoint, non-empty cycles in permutations of *n* objects, and since integer entropy is defined in terms of the expected value of the number of cycles over the set of permutations, it also has a clear combinatorial meaning. Since all finite groups are isomorphic to subgroups of the Symmetric group, every finite group has a corresponding "information functional", analogous to the Shannon entropy and a "number series", analogous to the harmonic numbers. The Cameron-Semeraro cycle polynomial is used to analyze the integer entropy for finite groups, and to characterize the series analogous to the Harmonic numbers. We introduce and use a reciprocal polynomial, the transposition polynomial that provides an additional tool and new insights. Broken symmetries and conserved quantities are linked through the cycle and transposition properties of the groups, and can be used to generalize the analysis of stochastic processes.

> *If you understand something in only one way, then you don't really understand it at all. The secret of what anything means to us depends on how we've connected it to all other things we know… that's what we mean by thinking!*
> — *Marvin Minsky*

## I.  Introduction

Information theory is focused on measures and arguments related to the transmission and quantitation of order and disorder, and these measures have been used to describe complexity as well. It has always been clear that since order and disorder are related to structure and symmetry that there must be intimate relations among the ideas of information and symmetry, but elucidation of fundamental connections has been sparse. There have also been many suggestions for the generalization of the Boltzmann-Gibbs-Shannon entropy [1-8]. These entropy formulas are often proposed based on an axiomatic approach, like that of Shannon or Khinchin [6-8]. One approach modifies the axioms, for example, abandoning the additivity axiom which leads to general and non-extensive entropies like the Tsallis or Renyi entropies [1,3], which have been connected to physical phenomena. Other approaches are based on physical assumptions like those of thermal equilibrium and processes active in a physical system [3,5,9,10], but there has been little mathematical connection between symmetries and information-entropy from a basic perspective that we know of. While there have been relations between entropy and groups previously proposed, they have been based on completely different principles [32-36] than proposed here.

Here we describe a simple and fundamentally new way (to our knowledge) of thinking about the entropy-symmetry problem and generalizing entropy in a fundamental way. We first connect the theory of finite groups, through specific sets of permutations, to information-entropy measures of the Shannon type. These measures are at the foundations of many areas of communications, complexity and randomness, statistical physics, and data analysis. We then use this connection to propose a natural mathematical generalization, which provides a new connection between information and symmetries.

There are many definitions of entropies, based on specific uses in physics, that have a range of properties. Breaking from the Boltzmann and Gibbs precepts, Tsallis first defined a new kind of entropy that is not extensive in the usual physical cases [3], but it is now clear that no entropy is actually extensive in all cases, and that the so-called Shannon-Khinchin axioms can be used with slight modifications to define a large range of useful functions. We are aware of these important extensions, but will use the Shannon definition here, closest to the Boltzmann-Gibbs structures, for our more fundamental generalization. Several complexity measures based on information have been suggested but most of these, including our own previous work [11-14], depend on uses of the Shannon information measure for single and multiple variables. The connection we make here with group symmetries may elucidate these complexity measure ideas as well, but we do not extend our reach to these in the present paper.

All Shannon-based information measures can be expressed as sums and differences of entropy functions, which includes multivariable functions like mutual information, complete correlation or multi-information, interaction information, and others. It includes the full range of measures expressible as Möbius inversions of entropies [14]. The estimation of entropies from data sets is a basic practical problem and is therefore central to general data analysis using information theory methods as well as in statistical physics. The distinction between entropy and information can be important but is not essential to the connections we focus on in this paper.

Since our general approach to connecting symmetry and information is based on group theory we comment briefly on its fundamental importance. The finite groups represent the bricks of a large and elegant mathematical edifice of enormous scope and importance. One of the great accomplishments of $20^{th}$ century mathematics is the "classification theorem" that defines the number and character of these bricks, some of which are simple, some mysterious and huge, but all of which are part of a structure with a clear, if remarkably subtle, architecture. All the deep and hidden meanings resident in these groups are far from evident. Since, by Caley's theorem, every finite group is isomorphic to a subgroup of a Symmetric group (the full range of permutations of labelled objects) we encompass, in principle, all the finite groups by specifying subgroups of the Symmetric group. Our approach does exactly this – define a prescription for an entropy or information functional for every finite group.

**Summary of Results**

We examine subgroups of permutations with the idea that the cycle structure of the permuted objects is the key to connecting the group symmetries to the general concept of entropy. The concept of the *integer entropy* is then shown to correspond to the Shannon definitions in specific limits of large numbers. Entropy, or information, is thus extended to a more complex architecture than the Shannon formulation, those reflected in the architectures of the finite groups. We take particular note of the fact that the choice of subgroup is, in itself, a specification of the breaking of the full permutation symmetry, so that the resulting entropy functional reflects the additional order of the subgroup.

The major results of this paper are summarized here in brief.

1. We define a measure of the partitions of $N$ objects, $\{n_i\}$, based on the number of cycles induced by the permutations of the subsets, which we call the "integer entropy." We show that this measure converges to the classical Shannon entropy in the double limits of large

*N* and the full Symmetric groups. Shannon entropy reflects "full permutation symmetry," which is broken by choosing a subgroup.

2. We identify a polynomial based on the rising factorial polynomial, determined by the partition $\{n_i\}$, that yields the integer entropy for the full Symmetric groups when evaluated at $x = 1$, and yields differentially weighted permutation cycles for other values of $x$. This polynomial is an instance for the Symmetric group of the cycle polynomial, defined by Cameron and Semeraro [28].

3. We extend this relation to subgroups of the Symmetric group, $G_N \subseteq S_N$, using the cycle polynomial, and show that the logarithmic derivative of the cycle polynomial for group $G_N$ evaluated at 1 is the equivalent for $G_N$, to the harmonic number for the Symmetric group. These are the expected cycle numbers for the group.

4. We show that the expected cycle number can be expressed as the sum of the reciprocals of the roots of the cycle polynomial.

5. Using these results, we determine the bounds for the coefficients of the cycle polynomial, and of the expected cycle number of a finite group.

6. Following Cameron and Semeraro we define a transposition polynomial using the minimum number of transpositions required to undo a permutation. This polynomial is reciprocal to the cycle polynomial and we show how its roots can be used similarly.

7. We suggest further that since the Symmetric group defines the usual integer entropy (the familiar Shannon entropy in the limit) and each finite group has a parallel integer entropy, that these "constrained" entropies (or "broken symmetry" entropies) can be viewed as "conditional integer entropies" in the sense that they are conditioned on the symmetry.

In an appendix of the paper, we illustrate the integer entropies for some specific subgroups, including the Alternating group, Cyclic group, and Dihedral group, and examine some of their properties.

## II. The Integer Entropy and Permutation Cycles

**Preliminaries**

If *G* is a finite group, let *G* be the permutation subgroup isomorphic to *G* for *N* labelled objects.

The Shannon entropy, $I_N(\{n_i\})$, for a set of discrete variables, $\{X_i\}$, with values $\{n_i\}$, where $\sum_{i=1}^m n_i = N$, is defined as

$$I_N(\{n_i\}) = -\sum_i \frac{n_i}{N} \log \frac{n_i}{N}$$

(1)

Here we have defined the probabilities associated with the numbers as the ratios of $\{n_i\}$ and $N$. If the number of cycles induced by the group element $\pi$, is $c(\pi)$, the average number of cycles for the group $G_N$ is

$$\langle C \rangle_{G_N} = \frac{1}{|G|} \sum_{\pi \in G} c(\pi),$$

(2)

**Definition:** we define the *integer entropy* as

$$J_{G_N} \equiv -\frac{1}{N} \sum_{i=1}^m n_i \left[ \langle C \rangle_{G_{n_i}} - \langle C \rangle_{G_N} \right] = \langle C \rangle_{G_N} - \frac{1}{N} \sum_{i=1}^m n_i \langle C \rangle_{G_{n_i}}$$

(3)

In this expression the entropy is defined, in the familiar fashion, as the expectation of the "point entropy" or "surprisal", which in our case is $\langle C \rangle_{G_N} - \langle C \rangle_{G_{n_i}}$. With this definition can then state our central result.

**Theorem 1**: The double limit of the <u>integer entropy</u>, which means expanding the group elements to include all permutations, $\lim_{G_N \to S_N}$, then increasing the number $N$ of permutated objects of the group, is equal to the <u>Shannon entropy</u> of $\{n_i\}$,

$$\lim_{N \to \infty} \left( \lim_{G_N \to S_N} (J_{G_N}) \right) = I(\{n_i\}),$$

(4)

where $S_N$ is the symmetric group, $I(\{n_i\})$ *is* the Shannon entropy, or information, of the partition $\{n_i\}$.

**Proof:** First, we expand the group $G_N$ so that it encompasses the full range of permutations, the Symmetric group. Then by the definition of the average number of cycles over the group of all permutations, $G \to S$ leads to

$$\langle C \rangle_{G_N} = \frac{1}{|G|} \sum_{\pi \in G} c(\pi) \Rightarrow \frac{1}{N!} \sum_{\pi \in S} c(\pi)$$

(5)

The average number of cycles of the full range of permutations of $N$ can be expressed in terms of the signless Stirling numbers of the first kind [18-20]. Using the known identity (presented, for example, in Benjamin and Quinn [19]),

$$\sum_{k=0}^{n} \begin{bmatrix} n \\ k \end{bmatrix} k = \begin{bmatrix} n+1 \\ 2 \end{bmatrix}$$

(6)

and recognizing that the signless Stirling number $\begin{bmatrix} n \\ k \end{bmatrix}$ counts the number of permutations of $n$ objects with $k$, non-empty, disjoint cycles, the summed expression is the number over all $n!$ permutations of the symmetric group, we have

$$\langle C \rangle_{G_n} = \sum_{k=0}^{n} \begin{bmatrix} n \\ k \end{bmatrix} \frac{k}{n!} = \frac{\begin{bmatrix} n+1 \\ 2 \end{bmatrix}}{n!} = H_n$$

(7)

since $\frac{\begin{bmatrix} n+1 \\ 2 \end{bmatrix}}{n!}$, is just $H_n$, the $n$th harmonic number [17,18]. Note that there are several other, well-known proofs of the harmonic number equivalence to the average number of permutation cycles. From Equation 7 it is clear that the expression for $J$ under the full symmetric group, is

$$J_S \equiv H_N - \frac{1}{N} \sum_{i=1}^{m} n_i H_{n_i}.$$

(8)

Since $\lim_{N \to \infty} (H_N - \ln(N)) = \gamma$, where $\gamma$ is the Euler-Mascheroni constant, we have

$$\lim_{N \to \infty} J_{G \to S} = \ln(N) - \frac{1}{N} \sum_{i=1}^{m} n_i \ln(n_i) = S(\{n_i\})$$

(9)

Because the fractions are equal to the probabilities, $\frac{n_i}{N} = p_i$, and the harmonic numbers converge to logarithms so that, finally, $S(\{n_i\}) = -\langle p_i \log p_i \rangle$ which concludes the proof.

An immediate implication of this theorem is that <u>every finite group</u> has its corresponding number series and integer entropy associated with it. There is always an expected number of cycles over any subset of permutations. Furthermore, all of them converge to the harmonic number version in the limits of large $N$ and the full Symmetric group. The appearance of the number of cycles induced has a strongly intuitive meaning here. The objects in a cycle maintain their order with respect to one another, thus the number of cycles is a kind of indication of disorder. In Equation 3 the difference between the two term on the right is the measure of the disorder induced in the entire set minus that in the partitions. The partition that minimizes the latter term maximizes the integer entropy.

### III. The Cycle Polynomial and the Integer Entropy

The cycle polynomial for the finite group $G_n$ is the generating function for a number of measures of the group and is particularly useful in calculating cycle properties. If $G_n \subset S_n$, and $c(\pi)$ is the number of cycles induced by the group element $\pi$, the cycle polynomial for $G_n$, as defined by Cameron and Semeraro [28], is

$$P_{G_n}(x) = \sum_{\pi \in G_n} x^{c(\pi)} = \sum_{k=1}^{n} g_{n,k} x^k$$

(10)

**Theorem 2:** If $G_n \subset S_n$, the upper bounds of the cycle polynomial coefficients are the corresponding signless Stirling numbers of the first kind, and the lower bound is zero.

$0 \leq g_{n,k} < \begin{bmatrix} n \\ k \end{bmatrix}$, for all $k$.

**Proof:** The sum over the elements of the Symmetric group can be separated into a sum over the subgroup elements and the disjoint set of elements in the complement of $G_n$, $\pi \in (S_n - G_n)$. Since all the terms are positive for $x > 0$, and

$$\sum_{\pi \in G} \delta_{c(\pi),k} = g_{n,k}$$

(11)

With the delta being the discrete Kronecker function, the proof is then direct.

**Corollary:** Since the number of group elements that induce $k$ cycles is $g_{n,k}$, the sum of these coefficients is obviously equal to $|G|$.

$$\sum_{k=1}^{n} g_{n,k} = \sum_{k=1}^{n} \sum_{\pi \in G} \delta_{c(\pi),k} = |G|$$

(12)

and the probability that a randomly chosen element of $G$ induces $k$ cycles is $\frac{g_{n,k}}{|G|}$.

**Theorem 3:** For any group $G_n \subseteq S_n$ the logarithmic derivative of the cycle polynomial is the average number of cycles for the group

$$\frac{1}{P_G(x)} \frac{\partial}{\partial x} P_G(x) \bigg|_{x=1} = \langle C \rangle_G$$

(13)

**Proof:** Since it is clear that $P_G(1) = |G|$, and that the derivative brings down $k$ in the form of the polynomial with $x^k$ components, and the proof is direct.

We can now express the entropy in terms of the cycle polynomial by using the logarithmic derivative[1] of $P_G(x)$, which we denote as $\check{P}_G(x)$, evaluated at $x=1$ for groups $G_n$. The group in its permutation representation acts on a set of size $N$, which is partitioned into $\{n_i\}$. The functional integer entropy, $J_G$, can be written as

$$J_G(\{n_i\}) = \langle C(N) \rangle_G - \sum_i n_i \langle C(n_i) \rangle_G$$

(14a)

and becomes

$$J_G(\{n_i\}) = \check{P}_{G_N}(1) - \sum_i n_i \check{P}_{G_{n_i}}(1)$$

(14b)

**Corollary:** Writing out the expressions for the average cycle numbers of $G_n$ and $S_n$, and applying Theorem 2 leads to the immediately to the bound $\langle C \rangle_G \leq \frac{H_n n!}{|G|}$, where the equality obtains for $G = S_n$.

---

[1] Note that the logarithmic derivative is the equivalent of the Cauchy transform of the cycle polynomial.

We can use the roots of the cycle polynomial now to express the expected cycle number in a different way.

**Theorem 4:** For any group $G_n \subseteq S_n$, if the roots of the cycle polynomial of $G_n$ are $\{r_k\}$ the average number of cycles is

$$\langle C \rangle_G = \sum_{k=1}^{N}(1-r_k)^{-1}$$

(15)

**Proof:** Theorem 3 gives the expected cycle number for $G_n$ when applied to the polynomial written as a product,

$$P_G(x) = \sum_{\pi \in G} x^{C(\pi)} = \prod_i (x-r_i)$$

(16)

$$\langle C \rangle_{G_n} = \check{P}_{G_n}(1) = \prod_i (1-r_i)^{-1} \sum_j \prod_{k \neq j}(1-r_k) = \sum_j (1-r_j)^{-1}$$

(17)

The range of the sum is over all roots, and there are $N$ roots for permutations on a set of $N$. Since the roots can only be zero, negative reals, or complex, the denominators are never zero (1 is never a root). Since we are counting fixed points and the identity is an element of $G_n$, the maximum number of cycles is $n$. This proves the theorem.

The coefficients of the polynomial, $g_{n,k}$, expressed in terms of the roots are then

$$g_{n,k} = \sum_{k\ tuples} \prod_{i=1}^{k}(-r_i), \qquad g_{n,n}=1$$

(18)

**Corollary:** Since by the Theorem 4 and the previous corollary we have the inequality

$$\sum_{k=1}^{n}(1-r_k)^{-1} \leq \frac{H_n n!}{|G|}$$

(19)

where $\{r_k\}$ are the roots of the cycle polynomial of $G_n$. Using the sum expression for the Harmonic numbers, we can insert a weighting factor into each term of the sum, $w_n(k)$, to make this an equality. This defines these weighting factors.

$$\sum_{k=1}^{n}(1-r_k)^{-1} = \sum_{k=1}^{n} \frac{w_n(k)}{k} = \langle C \rangle_{G_n}$$

(20)

Thus, the analog of the harmonic numbers for group $G_n$ can be written as a weighted form of the Harmonic number sum, the factor $\frac{n!}{|G_n|}$, being included in the $w_n(k)$. The weights are determined by the roots of the cycle polynomial:

$$w_n(k) = \left(\frac{k}{1-r_k}\right)$$

(21)

When the group is the Symmetric group the roots are $\{0,-1,-2,-3,\ldots n-1\}$, the polynomial is the rising factorial, the number of elements $|G_n| = n!$, and the $w_n(k)$ are equal to 1 for all $k$.

## IV. <u>**Counting Cycles Over Subsets of Permutations**</u>

In describing the properties of a subgroup, two kinds of weights describing the deviations from the Symmetric group are possible: in the sum over cycle lengths as in Equation 22.

The first sum over $k$ is a sum over fractions of cycles of *length k,* while the second is a sum over *numbers* of cycles, which depends on $n$ through the weights $y_n(k)$, which can be defined as weights on the signless Stirling numbers of the first kind, $\begin{bmatrix}n\\k\end{bmatrix}$. Both sums yield the expectation values of cycle numbers over permutations of $n$, as weighted, but the weightings are clearly distinct.

$$\langle C \rangle_{G_n} = \sum_{k=1}^{n} w_n(k)\frac{1}{k} = \frac{1}{n!}\sum_{k=1}^{n} y_n(k) \begin{bmatrix}n\\k\end{bmatrix} k$$

(22)

The signless Stirling number $\begin{bmatrix}n\\k\end{bmatrix}$ counts the number of permutations of $n$ objects having $k$ cycles for the Symmetric group, so the weights, $y_n(k)$, describe the deviations from the rising factorial.

The limiting case, with $w_n(k) = y_n(k) = 1$, the group is the full Symmetric group. We will address elsewhere the general question of what pairs of these distinct weighting functions generate identical series as determined by each specific subgroup.

The weightings in the sums over cycle lengths provides a tool for elucidating the relations between the cycles of the Symmetric groups and their subgroups. We can express the Harmonic numbers by using polynomials that provide a succinct expression for the weighted cycle numbers. With this we can calculate the cycle numbers induced by permutation subsets. First, a few definitions.

The rising factorial power of $x$, indicated by the notation $x^{(n)}$, is the polynomial defined by the product

$$x^{(n)} = x(x+1)(x+2)\ldots(x+n-1) \tag{23}$$

or equivalently, $\dfrac{x^{(n)}}{n!} = \binom{x+n-1}{n}$, as a formal expression. This is often called the Pochhammer symbol. The signless Stirling numbers of the first kind, used above, are the coefficients of this polynomial (for example, see [18]).

$$x^{(n)} = n!\binom{x+n-1}{n} = \sum_{k=0}^{n} \begin{bmatrix} n \\ k \end{bmatrix} x^k$$

(24)

As pointed out above, the rising factorial is actually the cycle polynomial of the Symmetric group. What we call the cycle function is the logarithmic derivative, $\check{P}_{G_n}(x)$.

Using the expression for the cycle polynomial in term of its roots, we see immediately that

$$\check{P}_{G_n}(x) = \frac{1}{P_{G_n}(x)} \sum_{k=1}^{n} \frac{P_{G_n}(x)}{x - r_k} = \sum_{k=1}^{n} \frac{1}{x - r_k}$$

(25)

where $\{r_k\}$ are the cycle polynomial roots.

## V. **The Transposition Picture and the Transposition Polynomial**

Permutations can be decomposed into transpositions (swapping two objects of the permuted set), and therefore the inverse can also be so decomposed. Given a permutation operator we can usefully think about the number of transpositions required to reorder the permuted set, restoring its original order. Since a result from Caley in 1849 relates the transposition number to the cycle number, we can use the cycle polynomial and introduce a new polynomial, the transposition polynomial to explicate these relations further. We show here that these are reciprocal polynomials. The idea of characterizing groups by their patterns of cycles and the pattern of minimum numbers of reordering transposition leads quite naturally for us to define a class of groups for which these are equal. We define and characterize some properties these symmetric, *balanced groups* and note some of their specific characteristics.

We begin with the relationship between cycle numbers of a given permutation the transposition decomposition due to Caley (1849). Every permutation can be decomposed into a set of disjoint cycles, such that the sum of their lengths is the number of objects permuted, if we count fixed points as cycles of length one. For a given permutation, $\pi$, of $n$ objects there are $C_n(\pi)$ cycles. Consider a single cycle of length $k$ in a given permutation. The cycle can be reversed; that is, put back into the original order, by a minimum of $k$-1 transpositions. If the objects in the cycle are ordered in the usual way we can see that beginning with the smallest label and applying $k$-1 transpositions we can obtain the original label order. Putting all the cycles back into the original order, and thereby undoing the permutation $\pi$, then requires only $n - C_n(\pi)$ transpositions. This gives the relationship between the cycle number and the minimum number of "reordering" transpositions, we call $T_n(\pi)$:

$$T_n(\pi) = n - C_n(\pi) \qquad (26)$$

As noted by Diaconis and Graham [32], this result was first obtained by Cayley in 1849 [33]. It is evident that the more cycles induced by a given permutation the fewer transpositions are needed to undo it. In the limiting case of all fixed points no transpositions are required. The ordered product of the undoing transpositions, $\{t_1, t_2, ... t_n\}$, is the inverse of the original permutation $\pi$. Keep in mind that the actions of transpositions on disjoint cycles is independent of order, but the order within each cycle is critical. Also note that because of the independent character of the disjoint cycles there are several possible orders of the full set of transpositions, even though they all constitute the same inverse of the permutation. The number of cycles generally indicates disorder and the transposition number, indicates order. Note that for fixed points this is not true, however, but it is for all cycles of length greater than one.

Using Equation 26 we can use the cycle polynomial to obtain an expression for the expected transposition number. "Expected" means the average over the group elements. The following theorem makes the connection between the average cycle number and the average transposition number.

**Theorem 5:**

For a permutation group on $n$ objects, $G_n$, the average, of the minimum number of transpositions need to reorder a permutation $\pi \in G_n$, over all the permutation elements, $T(\pi)$, is

$$\langle T \rangle_G = \sum_{k=1}^{n-1} \frac{-r_k}{1-r_k}$$

(27)

where $\{r_k\}$ are the roots of the cycle polynomial of $G_n$.

***Proof***: Given the relationship between minimum transposition numbers and cycle numbers we can write the cycle polynomial as:

$$P_G(x) = \sum_{\pi \in G} x^{n-T(\pi)} = x^n \sum_{\pi \in G} x^{-T(\pi)}$$

(28)

Expressing the polynomial in terms of its roots we have

$$P_G(x) = \prod_{k=1}^{n}(x - r_k)$$

(29)

and therefore

$$P_G(x) = x^n \prod_{k=1}^{n} \frac{(x-r_k)}{x} = x^n \prod_{k=1}^{n}\left(1 - \frac{r_k}{x}\right) \equiv x^n W_G(x)$$

(30)

Since the average cycle number over the elements of $G_n$ is the logarithmic derivative of $P_G(x)$ evaluated at $x = 1$, we have:

$$\check{P}_G(x) = \frac{n}{x} + \check{W}_G(x)$$

(31)

This means then that the logarithmic derivative $\check{W}_G(x)$, evaluated at $x = 1$ must be minus the expected transposition number, $\check{W}_G(x)|_{x=1} = -\langle T \rangle_G$, and

$$\check{P}_G(x)|_{x=1} = n - \langle T \rangle_G$$

(32)

Carrying out calculation of the logarithmic derivative of $W_G(x)$ we have the desired result.

**The transposition polynomial**

Equation 31 makes it clear that the expected transposition number for permutation group, $G_n$, is the logarithmic derivative of $-W_G(x)$ evaluated at $x = 1$. If we define the transposition polynomial as

$$Q_G(x) = \sum_{\pi \in G} x^{T_n(\pi)}$$

(33)

then by the same arguments made for the cycle polynomial the logarithmic derivative at $x = 1$ is the expected transposition number. Thus $Q_G(x) = -W_G(x)$ and $\check{Q}_G(x)|_{x=1} = \langle T \rangle_G$. Note however, that this polynomial differs from the cycle polynomial in an important respect. Since the identity element generates $n$ cycles for a set of $n$ objects the cycle polynomial always has a term $x^n$, and no constant term, since there is always a cycle in any permutation. The transposition polynomial, on the other hand, will have no $x^n$ term, and always have a constant term.

**Reciprocal polynomials**

From Equation 28 we can see directly that a change of variables from x to 1/x in $P_G(x)$ yields $Q_G(x)$, and that they are closely related.

$$P_G\left(\frac{1}{x}\right) = x^{-n} \sum_{\pi \in G} x^{T(\pi)} = x^{-n} Q_G(x)$$

(34)

There is a dual relationship between the cycle and transposition polynomials, $P_G(x)$, and $Q_G(x)$. They are reciprocal, or reflected polynomials. The coefficients of these two polynomials are reversed, so that

$$P_G(x) = \sum_{k=1}^{n-1} g_k x^k \text{, and } Q_G(x) = \sum_{k=2}^{n} f_k x^k$$

(35)

where $f_k = g_{n-k}$.

From this relationship we can put bounds on the coefficients of the transposition polynomial.

**Proposition 3:** If $G_n \subset S_n$, the upper bounds of the transposition polynomial coefficients are the corresponding signless Stirling numbers of the first kind, and the lower bound is zero.

$0 \leq f_{n,k} < \begin{bmatrix} n \\ n-k \end{bmatrix}$, for all $k$.

Now the question of the relation between the roots of the transposition polynomial and the cycle polynomial is simply answered. Since $x^n P_G\left(\frac{1}{x}\right) = Q_G(x)$ we can write $Q_G$ in terms of the roots of $P_G$

$$Q_G(x) = \prod_{k=1}^{n} (1 - x r_k)$$

(36)

Thus the roots of the transposition polynomial are the reciprocals of those of the cycle polynomial. The transposition polynomial for the symmetric group, where $r_k = 1 - k$ is then

$$Q_{S_n}(x) = \prod_{k=2}^{n} (x(k-1) + 1) = (x+1) \cdot (2x+1) \cdot (3x+1) \ldots ((n-1)x + 1)$$

(37)

The roots are obviously $\{-1, -\frac{1}{2}, -\frac{1}{3}, -\frac{1}{4}, \ldots, -\frac{1}{n}\}$. Recall that the cycle polynomial for the symmetric group is the rising factorial, $x^{(n)}$. It is an interesting curiosity as well that the sum of the roots of the transposition polynomial for $S_n$ is the harmonic number, $-H_n$, which is the negative of the average number of cycles of the symmetric group.

$$H_n = \sum_{k=1}^{n} -q_k(S_n)$$

**Theorem 6:**

For a permutation group on a set of $n$, $G_n$, the average over all the permutation elements of $G_n$ of the minimum number of transpositions needed to reorder a permutation $\pi$, $T(\pi)$, is

$$\langle T \rangle_G = \sum_{k=1}^{n} \frac{1}{1 - q_k}$$

(39)

**Corollary:** The reciprocity between the cycle and transposition polynomials, leads directly to this result

$$\langle C \rangle_G = \sum_{k=1}^{n} \frac{-q_k}{1-q_k}$$

(40)

## V. <u>Discussion</u>

There have been many proposals for the generalization of the Boltzmann-Gibbs-Shannon entropy and information [1-8], and here we have proposed a new one, linking each of them specifically to a finite group. The major difference between this and previous generalizations is that we do not use an axiomatic approach, which keys on the axioms of Shannon or Khinchin [6-8], or a physics-related approach either, but rather an intuitive mathematical one. The justification for taking this approach, without grounding it in the physical or axiomatic interpretation of an entropy or information functional, is that the mathematical generalization itself gets to the heart of one of major issues in the uses of such functionals, namely the questions of the relations among symmetry, information, order, and disorder. The close connection to group theory and symmetry emphasizes this aspect and promises further insights with a focus on the details of symmetries and symmetry breaking. The symmetry here is represented by the group action on a set of labeled objects and the entropy measures reflect just those symmetries. It is clear that the probabilities are directly related to the number of objects being permutated, and the surprisals and entropies related in an indirect way through the group. The idea of broken symmetry is simply the removal of one or more of the group elements, so that any subgroup of the Symmetric group can be viewed as embodying broken symmetries. With each broken symmetry there is a corresponding conserved quantity, which we will discuss in future papers. The idea is briefly illustrated for the Dihedral group in the Appendix.

We have made the connection between constraints on the distribution of probabilities and the number series, which are generalizations of harmonic numbers that correspond to a given subgroup. Designating a specific group we can then calculate the corresponding series, which is the expected number of cycles for this subgroup of the Symmetric group. However, it is yet unclear how one can go from a distribution or number series to the corresponding group. It is clear that the key parameters are the cycles induced by the permutations, so the conjugacy classes of the groups will be central objects to this connection.

A natural question arises: Are there some useful series that are associated with no group at all, perhaps a semigroup or other subset of permutations that is neither? This should certainly be the case as there are only a finite set of "classes" of finite groups and there is no such constraint on the algebraic series. On the other hand, it is easy to prove that every finite group has a series.

Based on our results we propose an interesting way to think about comparing the integer entropy with the Shannon entropy; that is, purely in terms of the roots of the cycle polynomial. If we think of the probability of the $i$th element of a string as $\frac{n_i}{N}$, we can think of the analog of the log of this probability as

$$\langle C \rangle_{n_i} - \langle C \rangle_N = -\sum_{k=n_i}^{N} \frac{1}{1 - r_k}$$

(41a)

where $\{r_k\}$ are the roots of the cycle polynomial of the group, or in terms of the roots of the transposition polynomial, $\{q_k\}$.

$$\langle C \rangle_{n_i} - \langle C \rangle_N = \sum_{k=n_i}^{N} \frac{q_k}{1-q_k}$$

(41b)

For the Symmetric group, of course, this is the difference of the harmonic numbers, $H_{n_i} - H_N$, which converges to $\log \frac{n_i}{N}$ for large numbers. Note that while the roots can be complex, the sum is real (see appendix.)

It is intuitive to attribute maximal disorder in a set of $N$ objects to the result of the full symmetric group, $S_N$, acting on it, since the symmetric group contains all possible permutations of the set of $N$ objects. Subgroups with fewer permutations represent more order, and carving away some subset of permutations from the full Symmetric group to break the symmetry and impose some additional order may be interpreted as putting certain constraints on the distribution. Keep in mind, however, that the values analogous to $\log \frac{n_i}{N}$ are differences (analogous to the differences between harmonic numbers) and can actually be larger or smaller than their analogs. In Appendix A we illustrate this explicitly for several specific groups, including the Alternating group, Cyclic group, and the Dihedral group. The result for the cyclic groups points to an unexpected effect in that the expected number of cycles has a highly variable component since it depends on the number of divisors of the group order, and takes on some stochastic character even though the average trend increases logarithmically.

There are many possible partitions of $N$, $\{n_i\}$, ( $N = \sum_{i=1}^{m} n_i$), and the permutations act on both the partitioned sets and whole set to define the entropy functional. Consider further that all permutations have the effect of partitioning a set of integers into disjoint cycles, from fixed points to full length cycles, so we could also think of any partition of $N$ into $\{n_i\}$ as the result of a single permutation that yields the subsets in each disjoint cycle. Since all the elements of any conjugacy class of permutations induce the same cycle structure we can identify each distinct partition of $N$, $\{n_i\}$, with a specific conjugacy class. The connection of this problem with conjugacy theory will be interesting to explore further.

We have described here a fundamental link between the entropy and the information measures of Shannon and finite groups. Theorem 1 generalizes this entropy function in a new way, showing that the Shannon information is essentially the limiting case of a much richer, more complex mathematical structure. The fact that the entropy maxima can be reached for a variety of non-uniform partitions (see Alternating group example in Appendix A) strongly suggests that the fundamental and celebrated notion of maximizing entropy modulo various constraints, can naturally be extended beyond the usual Lagrange multiplier methods to account for constraints specified by the specific choice of group. It seems that the group symmetries can be extended for most purposes to group element subsets as constraints of system properties, and any subset can be used to generate subgroups by multiplication. We suggest that in the convergence of symmetries, the combinatorics tools of permutations, like the symbolic methods of generating functions, a broad and more powerful theory of complexity and information can be built using these ideas. In this paper we have introduced the basic idea, but much remains to be done, and many extensions are possible.

While we have taken a distinctly algebraic approach here, connecting an entropy measure to permutations and their subgroups allows us, in principle, to describe this in geometric terms by using the recently explored polytopes associated with these combinatorics problems, the permutahedrons [26]. Since the structure of relationships among the permutations involved in the integer entropies can be described in geometric terms, there may well be also a rich diversity of results in this exploration.

A number of authors have previously reported on connections between information theory and groups [33-38]. These connections have focused on distributions on groups [19], on the correspondence between the information inequalities and group inequalities, on the connections with group lattices, and on the connection between entropy and topology, the homological nature of entropy [35]. The definition of the Shannon information as a functional is the centerpiece of all these investigations. While it is likely that the results reported in these papers are actually reflections of some of the deeper connections between group theory and information theory that are relevant to our approach, these connections remain largely unexplored, and should be addressed in future work. One important class of questions to answered, for example, is this. Since the Shannon entropy corresponds to the symmetric group, what group corresponds to the Renyi entropy and other forms [1-8] ?

The specific connection of permutation cycles with entropy that we have focused on seems to open an important and novel way into the use of combinatorics and its mathematical apparatus for describing order and disorder and the measures of information in complex systems. There are many further connections provided by this link because the permutation cycle field is linked to analytic number theory [21, 22], and other intriguing connections are similarly suggested. It will be both challenging and exciting to push these ideas forward into actual applications in physics, in machine learning, inference, and in data analysis.

Acknowledgements: I am grateful for stimulating discussions with Nikita Sakhanenko and James Kunert-Graf, and comments on the manuscript by them and by Alex Skupin. Fan Chung provided sage and sobering advice on an early outline of this manuscript, for which I thank her gratefully, even if I haven't followed it fully. This work was supported in part by the Bill and Melinda Gates Foundation, the NIH (5U01HL126496), and the Pacific Northwest Research Institute. The central idea was conceived much earlier under partial support by the NSF (EAGER grant IIS-1340619).

# Appendix A. Number series analogs of Harmonic numbers for specific subgroups

## A.1 The Alternating group

The Alternating group of degree $n$, $A_n$, is the group of all elements that induce even permutations of $n$ objects. Even permutations are those that can be produced by an even number of transpositions (all elements can be decomposed into products of transpositions.) $A_n$ is a subgroup of $S_n$. Note that the odd permutations are not a subgroup since multiple odd permutations can be even so odd permutation elements cannot be closed, failing that group criterion. There are $n!/2$ elements of $A_n$, so it is a large subgroup of $S_n$. There are a number of well-known properties of $A_n$: for example, all $A_n$ for $n > 2$ are simple groups. It also has a number of properties in common with the symmetric group. Let's have a look at the cycle structure of $A_n$ first to compare the average cycle numbers for the Alternating group with the harmonic numbers. This result is relatively simple, if surprising. The number of cycles of each length for the first few $n$ of $S_n$ and $A_n$ are easy, to calculate and the table of the results up to $n=8$ is here, A1.

| Cycle type cycles>1 | 0 | 1 2 | 2 3 | 3 2,2 | 4 4 | 5 3,2 | 6 5 | 7 2,2,2 | 8 4,2 | 9 3,3 | 10 6 | 11 3,2,2 | 12 5,2 | 13 4,3 | 14 7 | 15 2,2,2,2 | 16 4,2,2 | 17 3,3,2 | 18 6,2 | 19 5,3 | 20 4,4 | 21 8 | Σ | Harmonics |
|---|---|---|---|---|---|---|---|---|---|---|---|---|---|---|---|---|---|---|---|---|---|---|---|---|
| No. Cycles | 1 | | | | | | | | | | | | | | | | | | | | | | | |
| 1 S | 1 | | | | | | | | | | | | | | | | | | | | | | 1 | 1.0000 |
| 1 A | 1 | | | | | | | | | | | | | | | | | | | | | | 1 | 1.0000 |
| No. Cycles | 2 | 1 | | | | | | | | | | | | | | | | | | | | | | |
| 2S | 1 | 1 | | | | | | | | | | | | | | | | | | | | | 2 | 1.5000 |
| 2A | 1 | 0 | | | | | | | | | | | | | | | | | | | | | 1 | 2.0000 |
| No. Cycles | 3 | 2 | 1 | | | | | | | | | | | | | | | | | | | | | |
| 3S | 1 | 3 | 2 | | | | | | | | | | | | | | | | | | | | 6 | 1.8333 |
| 3A | 1 | 0 | 2 | | | | | | | | | | | | | | | | | | | | 3 | 1.6667 |
| No. Cycles | 4 | 3 | 2 | 2 | 1 | | | | | | | | | | | | | | | | | | | |
| 4S | 1 | 6 | 8 | 3 | 6 | | | | | | | | | | | | | | | | | | 24 | 2.0833 |
| 4A | 1 | 0 | 8 | 3 | 0 | | | | | | | | | | | | | | | | | | 12 | 2.1667 |
| No. Cycles | 5 | 4 | 3 | 3 | 2 | 2 | 1 | | | | | | | | | | | | | | | | | |
| 5S | 1 | 10 | 20 | 15 | 30 | 20 | 24 | | | | | | | | | | | | | | | | 120 | 2.2833 |
| 5A | 1 | 0 | 20 | 15 | 0 | 0 | 24 | | | | | | | | | | | | | | | | 60 | 2.2333 |
| No. Cycles | 6 | 5 | 4 | 4 | 3 | 3 | 2 | 3 | 2 | 2 | 1 | | | | | | | | | | | | | |
| 6S | 1 | 15 | 40 | 45 | 90 | 120 | 144 | 15 | 90 | 40 | 120 | | | | | | | | | | | | 720 | 2.4500 |
| 6A | 1 | 0 | 40 | 45 | 0 | 0 | 144 | 0 | 90 | 40 | 0 | | | | | | | | | | | | 360 | 2.4833 |
| No. Cycles | 7 | 6 | 5 | 5 | 4 | 4 | 3 | 4 | 3 | 3 | 2 | 3 | 2 | 2 | 1 | | | | | | | | | |
| 7S | 1 | 21 | 70 | 105 | 210 | 420 | 504 | 105 | 630 | 280 | 840 | 210 | 504 | 420 | 720 | | | | | | | | 5040 | 2.5929 |
| 7A | 1 | 0 | 70 | 105 | 0 | 0 | 504 | 0 | 630 | 280 | 0 | 210 | 0 | 0 | 720 | | | | | | | | 2520 | 2.5690 |
| No. Cycles | 8 | 7 | 6 | 6 | 5 | 5 | 4 | 5 | 4 | 4 | 3 | 4 | 3 | 3 | 2 | 4 | 3 | 3 | 2 | 2 | 2 | 1 | | |
| 8S | 1 | 28 | 112 | 210 | 420 | 1120 | 1344 | 420 | 2520 | 1120 | 3360 | 1680 | 4032 | 3360 | 5760 | 105 | 1260 | 1120 | 3360 | 2688 | 1260 | 5040 | 40320 | 2.7179 |
| 8A | 1 | 0 | 112 | 210 | 0 | 0 | 1344 | 0 | 2520 | 1120 | 0 | 1680 | 0 | 0 | 5760 | 105 | 0 | 0 | 3360 | 2688 | 1260 | 0 | 20160 | 2.7357 |

**Table A1**. Table of the cycle numbers and lengths for $S_n$ and $A_n$. The cycle type is indicated at the top, and we examine the cycles for each $n$ as shown. 2S and 2A, for example, indicate the groups $S_2$ and $A_2$. The column marked "Harmonics" indicates the harmonic number for $S_n$ and that for $A_n$ just below it ("harmonics" for $A_n$ are the average number of cycles).

These "alternating harmonics" are calculated as the average number of cycles for a given $n$. The average number of cycles for each $n$ for the $A_n$ are clearly different from the harmonic numbers. What is also clear, however, is that as $n$ increases the two series rapidly converge. The profiles of the numbers as calculated from the cycle structure of the groups are shown in the plot below.

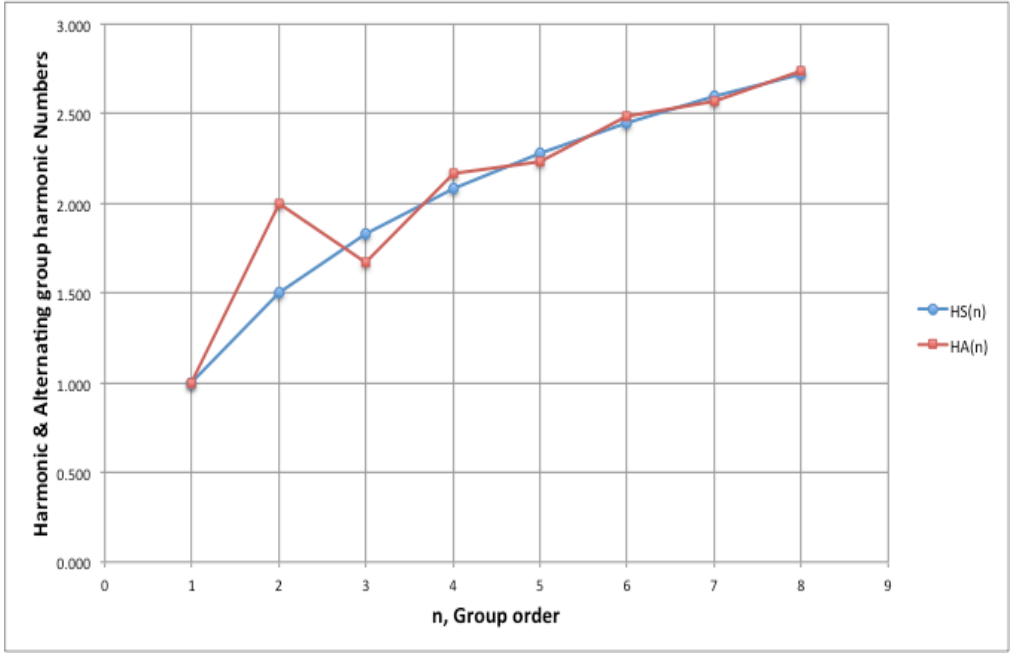

**Figure A1**. The "Alternating harmonic" and harmonic numbers for the first few integers. Harmonic numbers, as calculated from the cycle structure of the Symmetric group are shown in blue, and the same series for the Alternating group of the same degree are shown in red. The two converge rapidly as *n* increases.

It is surprising that the two series seem to converge very quickly since half of the permutations in $S_n$ are eliminated from the Alternating group for all *n*. Since we know that the harmonics are defined by a simple series, the partial sums of 1/k, the question arises as to what the summation expression is for the alternating version of the harmonics. Let's call these numbers the $\{A_n\}$, and now determine how to express the cycle structure number in a summation series. The answer to this question, it is easy to show using exponential generating functions that the series, to be compared with the harmonics in this case, is:

$$H_n = \sum_{k=1}^n \frac{1}{k}$$

$$\mathcal{A}_n = 2 + \sum_{k=3}^n \left(\frac{1}{k} + (-1)^k \frac{2}{k(k-2)}\right), \quad n > 2 \qquad (A1)$$

$$\mathcal{A}_n = 2 + \sum_{k=3}^n \left(\frac{1}{k}\right)\left(\frac{k - 2 + 2(-1)^k}{(k-2)}\right), \quad n > 2$$

$$\mathcal{A}_1 = 1, \qquad \mathcal{A}_2 = 2$$

The weighting factor defined in the main text, $w(k)$, in this case is therefore

$$w(k) = \left(\frac{k - 2 + 2(-1)^k}{(k - 2)}\right)$$

It is evident that $\lim_{n \to \infty} (\mathcal{A}_n - H_n) = 0$.

While it is clear that this subgroup is similar to the Symmetric group in one sense, and the number series converges quickly to the Harmonics, it is certainly not intuitively obvious. Neither is it obvious that this is common among the subgroups of the Symmetric group (all finite groups).

From [18] we have an expression for the cycle polynomial of the Alternating group, $\mathcal{A}_n$ (see Theorem 11).

$$2P_{\mathcal{A}_n}(x) = \prod_{k=0}^{n-1}(x - k) + \prod_{k=0}^{n-1}(x + k)$$

The difference between the harmonic numbers and the Alternating group number series implies the following. If the occupation of any state (the $\{n_i\}$) is not very low, all $n$'s are greater than 5, for example, the integer entropy function for the Alternating group is very close to the Symmetric group entropy, and close to the Shannon entropy. For low occupation numbers, small partition numbers, however, there will be a significant difference. A few examples show where the differences can arise. Direct calculation of the integer entropies $J_S$ and $J_A$ (Alternating group) for a total of $N=25$ objects and 5 partitions shows that both are maximal for a uniform partition: (5,5,5,5,5). The same is true for $N= 24$ and 28 for 4 partitions: uniform is maximal for both $J_S$ and $J_A$. However, for $N=16$ with 4 partitions, while $J_S$ is maximal for the uniform partition, $J_A$ is maximal for the non-uniform partition (3,3,5,5).

## A.2 The Cyclic group

The cyclic group is a very simple group in the sense of its symmetries and is Abelian. Taking our lead from the results in [28], we have for the cycle polynomial of the cyclic group on $n$ objects, where $\varphi$ is Euler's totient function,

$$P_{C_n}(x) = \sum_{d|n} \varphi(d) x^{n/d}$$

(A2)

which is the same as the dihedral group (next section) without the terms for the reflection symmetries.

The total number of cycles for the group $C_n$ is

$$\frac{\partial}{\partial x} P_{C_n}(x)|_{x=1} = \sum_{d|n} \frac{n}{d} \varphi(d)$$

(A3)

and since the total number of group elements is

$$|C_n| = \sum_{d|n} \varphi(d) = n$$

(A4)

the expected number of cycles is given by

$$\langle C \rangle_{C_n} = \sum_{d|n} \frac{\varphi(d)}{d}$$

(A5)

This is a rather complex function. It is clear that there is a big difference between this and the corresponding equation for the Harmonic numbers. There is no apparent recursion form of this relation, as the divisors of $n$ do not admit of such a relation. The cyclic groups have a progressive number of elements with $n$, but the number of divisors is certainly not progressive. We only have to look at the prime $n$'s to see this. For n=5, for example, there are 4 group elements with a cycle number of 1, and 1 group element with a cycle number of 5. For the next group, $n=6$ on the other hand, we have 1 element with cycle number 6, 1 element with a cycle number of 3, 2 with a cycle number of 2, and 2 with a cycle number of 1. The cycle polynomial for $C_5$ is $x^5 + 4x$, whose roots are $\{0, \sqrt{2}e^{i\frac{\pi}{4}}, \sqrt{2}e^{-i\frac{\pi}{4}}, \sqrt{2}e^{i\frac{3\pi}{4}}, \sqrt{2}e^{-i\frac{3\pi}{4}}\}$. The appearance of complex roots may be surprising so let us calculate the $\langle C \rangle_{C_5}$ to see how it sorts itself out. (It simplifies rapidly when we note that $\cos\left(\frac{\pi}{4}\right) = \sin\left(\frac{\pi}{4}\right) = \frac{1}{\sqrt{2}}$)

$$\langle C \rangle_{C_5} = 1 + \frac{1}{1 - \sqrt{2}e^{i\frac{\pi}{4}}} + \frac{1}{1 - \sqrt{2}e^{-i\frac{\pi}{4}}} + \frac{1}{1 + \sqrt{2}e^{i\frac{\pi}{4}}} + \frac{1}{1 + \sqrt{2}e^{-i\frac{\pi}{4}}}$$

$$\langle C \rangle_{C_5} = 1 + \frac{1 - \sqrt{2}e^{-i\frac{\pi}{4}}}{3 - 2\sqrt{2}\cos\frac{\pi}{4}} + \frac{1 - \sqrt{2}e^{i\frac{\pi}{4}}}{3 - 2\sqrt{2}\cos\frac{\pi}{4}} + \frac{1 + \sqrt{2}e^{-i\frac{\pi}{4}}}{3 + 2\sqrt{2}\cos\frac{\pi}{4}} + \frac{1 + \sqrt{2}e^{i\frac{\pi}{4}}}{3 + 2\sqrt{2}\cos\frac{\pi}{4}}$$

$$\langle C \rangle_{C_5} = 1 + \frac{2 + 2}{3 + 2} = 1.8$$

(A6)

which agrees with Equation A5. So even though the roots are complex, the imaginary parts of the roots cancel out in the sum to yield a real value as expected. Note that the requirement that the average cycle number be real means that if the roots are Complex, $\{r_i\} = \{R_i + I_i\}$, then the real and imaginary parts of the roots must be related such that

$$\sum_i \frac{I_i}{(1 - R_i)^2 + I_i^2} = 0$$

(A7)

The cycle polynomials for all prime $n$'s have a simple structure. If $p$ is prime, we have

$$P_{C_p}(x) = x^p + (p - 1)x$$

The roots of $P_{C_p}(x)$ are 0, and $\{e^{-i\frac{\pi}{p-1}(1+2k)} \setminus k \in (0 \text{ to } \frac{p+1}{2})\}$. Since we have derived two different expressions for the $\langle C \rangle_{C_n}$ they must be equal. Putting these arguments together then we have proven:

**Proposition:** For the cyclic groups $\langle C \rangle_{C_p}$ where $p$ is prime

$$\langle C \rangle_{C_p} = \check{P}_{C_p}(1) = \sum_{j=1}^{N}(1-r_j)^{-1} = \frac{2p-1}{p} = \sum_{k=0}^{\frac{p+1}{2}}\left[\frac{1}{1-e^{-i\frac{\pi}{p-1}(1+2k)}}\right]$$

(A8)

The closed form expressions for non-prime orders are more complex but clear. Our calculations of $\langle C \rangle$ for some other, non-prime $n$'s yield similar results, confirming our conclusion.

In addition, it is clear from Equation 21 in the main text that that the cycle function for the prime number cyclic groups are

$$\check{P}_{C_p}(x) = \sum_{k=0}^{\frac{p+1}{2}}\left[\frac{1}{1-e^{-i\frac{\pi}{p-1}(1+2k)}}\right]$$

(A9)

It is clear from the above that the limit of the sum over the roots for prime $p$ is an integral constant, namely 2,

$$\lim_{p\to\infty}\sum_{k=0}^{\frac{p+1}{2}}\left[\frac{1}{1-e^{-i\frac{\pi}{p-1}(1+2k)}}\right] = 2$$

(A10)

To illustrate the overall trend for the number series we calculate the values of $\langle C \rangle_{C_n}$ for the first few $n$, and the result is shown in Figure A2. There is a great deal of scatter caused by the occurrence of the totient function and the number of divisors. It may be something of a surprise that an increase in $n$ can lead to a decrease in $\langle C \rangle_{C_n}$, for example.

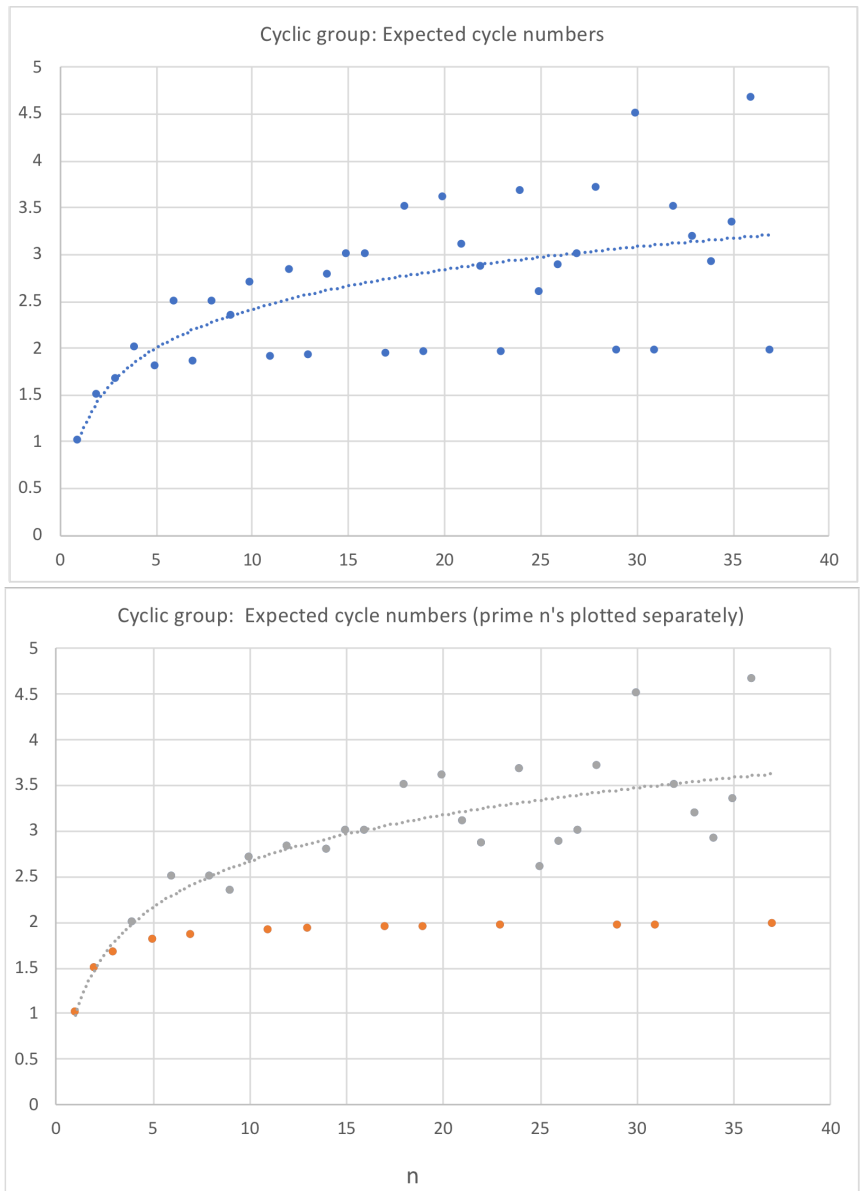

**Figure A2**. The Expected cycle numbers for the cyclic groups, $C_n$. The upper panel shows the full set of *n*'s together (with a fit to a logarithmic function), while the lower separates out the prime *n* series (the orange points), with a fit to the logarithmic function excluding the prime *n*'s. These latter converge rapidly to 2 as the number increases, while the non-prime values increase logarithmically with some significant scatter.

It is evident that the scatter arising from the pattern of primes and divisors is nonetheless following a logarithmic pattern, as further calculation shows. The primes form a pattern of their own, distinct from the non-prime *n*, and the average number of cycles converges quickly to 2 for the primes as they increase without bound. Since the symmetric group has a simple sum of reciprocals that converge to a logarithmic form, the harmonic numbers, it is interesting to compare the series $\langle C \rangle_{C_n}$ to that of $\langle C \rangle_{S_n}$. In Figure A3 we show the comparison between them – the harmonic numbers rise significantly faster with *n*.

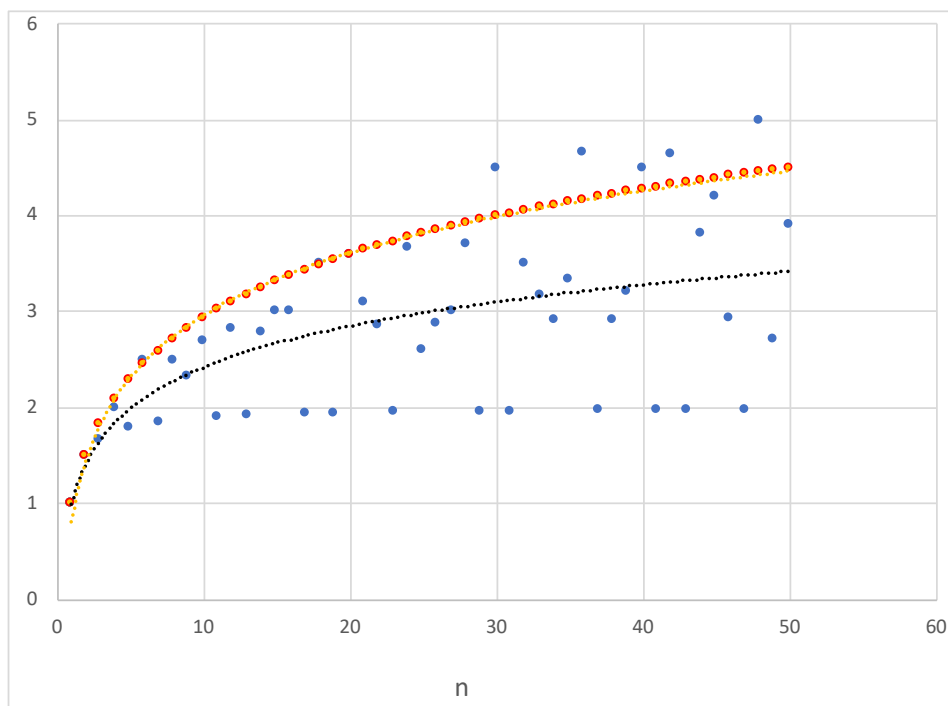

**Figure A3**. A comparison of the logarithmic trends of the harmonic numbers ($S_n$), shown as the orange points, with the corresponding numbers for the cyclic groups ($C_n$), the blue points. The black dotted curve is the logarithmic least squares fit to the cyclic group cycle averages.

### A.3 The Dihedral group

The dihedral group is one of the simplest non-Abelian groups, making it relatively easy to work out the CP since it can be viewed as an extension of the cyclic groups. It is the group of symmetries of the regular polygons with n sides. Designated $D_n$, it includes rotations, and reflections through all symmetry axes. In terms of permutations, if we track the numbers of cycles induced by each permutation element, we can map it into a permutation group and derive the cyclic polynomial. The total number of permutation elements in the group $D_n$ is 2*n*, a much smaller number than the *n*! elements of the full $S_n$. The pattern of symmetries and cycles is different for *n* odd or even. A table of the numbers of cycles and permutations inducing these cycle structures illustrates the pattern by showing the number of permutations of $D_n$ inducing the specific number of cycles (reflecting the symmetry of the *n* polygon).

___

<u>Odd *n*:</u>

| Cycle structure | All points fixed | Fixed points & transpositions | Cycles (for all *d*) |
|---|---|---|---|
| Number of Cycles | $n$ | $\frac{n-1}{2}$ | $n/d$ |
| Number of Permutations | 1 | $n$ | $\varphi(d)$ |

<u>Even *n*:</u>

| Cycle structure | All points fixed | 2 fixed points | All transpositions | Cycles (for all *d*) |
|---|---|---|---|---|
| Number of Cycles | $n$ | $\frac{n-2}{2}$ | $\frac{n}{2}$ | $n/d$ |
| Number of Permutations | 1 | $\frac{n}{2}$ | $\frac{n}{2}$ | $\varphi(d)$ |

___

**Table A2**. Cycle structure information for the dihedral groups, where the *d* are the number of divisors of *n*, and $\varphi(d)$ is Euler's totient function.

Here $\varphi$ is Euler's totient function, the function that counts the positive integers up to *d* that are relatively prime to *n*. From this table we can simply write out the polynomial using the definition of CP.

$$\text{n odd}: P_{D_n}(x) = x^n + nx^{\frac{n-1}{2}} + \sum_{d|n} \varphi(d) x^{n/d}$$

$$\text{n even}: P_{D_n}(x) = x^n + \frac{n}{2}x^{\frac{n}{2}} + \frac{n}{2}x^{\frac{n-2}{2}} + \sum_{d|n} \varphi(d) x^{n/d}$$

(A11)

The number of elements of the group, $P_{D_n}(1)$, is $2n+1$ for both odd and even *n*. The closed form expression for the expected cycle number can be easily calculated using the results from the previous section, but the calculation of the roots of (A10) is a bit more complex.

**<u>Symmetries and order parameters</u>**

The dihedral group comparison provides us with a simple example of the role of symmetries. It is useful to express this notion in the language common to physicists: what symmetries are broken, and what orderings are driven by the broken symmetries? The symmetries - that is the group elements (permutations) that the symmetric group prescribes - is many more than those of the subgroups considered here, and the dihedral group in particular. The broken symmetries for the dihedral group are the collection of elements of the symmetric group that are not elements of the dihedral group – the group complement. If we think of the occupation of a state by *n*-objects, the objects are fully interchangeable and indistinguishable under the symmetric group. However, under the dihedral group they can be viewed as the nodes in an *n*-polygon, for which all reflection and rotations are permitted, but not other node permutations. If we view any set of objects as an *n*-polygon there are three classes of possible permutations to consider: reflections, rotations, and the others which we can refer to as "twists." For the dihedral group, relative to the symmetric group the twist symmetries are broken, and the rotations and reflections are retained. This can be visualized for *n*= 4, and n=6 as shown in Figure A4.

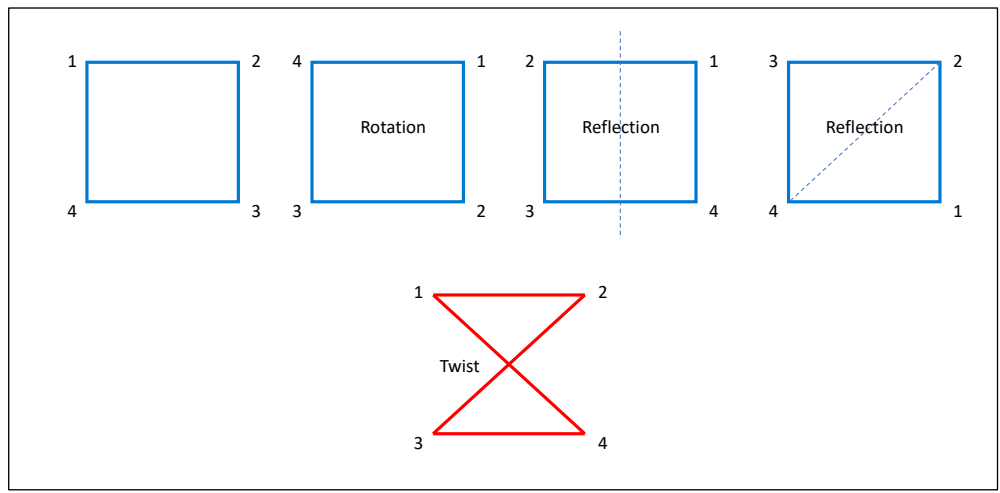

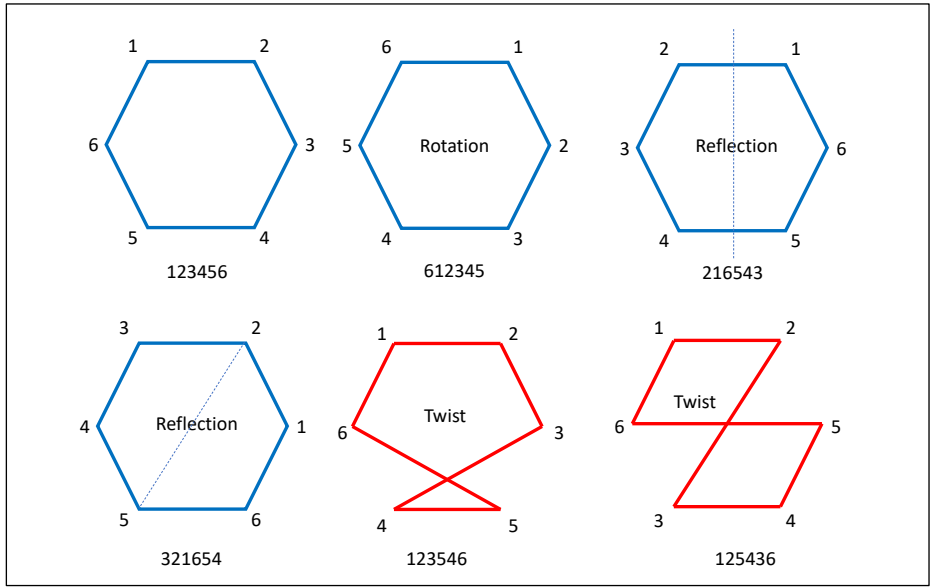

**Figure A4**. Permutation class examples for $n=4$ (**a**) and $n=6$ (**b**) as seen in terms of an $n$=polygon. The blue figures represent the dihedral group permutations, while the red, "twist," permutations represent those not in $D_6$. The permutation states are indicated under each figure.

It is relatively easy to see the order parameter for the broken symmetry called "twist." An order parameter is a quantity that is conserved by the subgroup, but not by the full group – it is not conserved by the broken symmetry, and therefore the broken symmetry and the order parameter are linked. One way of defining the order parameter quantitatively might be the relative area of the polygon if one orders the objects and maintains the edges between adjacent objects. The full permutation group does not conserve this area, while the dihedral group does, as shown in Figure A4. The broken symmetry and the corresponding order parameter are clear in this case. The comparison of the two symmetries, $D_n$ and $S_n$, is therefore somewhat like comparing two phases before and after a phase transition, where the polygon area is the order parameter of the ordered phase. In this case the area is a function of the occupation number of the state, $n$.

Note that neither the cyclic or dihedral groups are normal groups. This is evident from the fact that they do not consist of the unions of full conjugacy classes. A quick look at the elements of $C_4$ and $D_4$ makes this point. Consider, for example, the conjugacy class of cycles of 4 elements, where it is evident that there are some elements not in $C_4$ and $D_4$, and some elements that are in.

## Appendix B. Balanced groups

Considering the reciprocity between the cycles and the transpositions of permutations it is natural to define a kind of group that exhibits them in equal numbers. We therefore define *balanced groups*.

**Definition**: We define a *balanced group* as a group for which the average cycle number, the number of cycles induced, averaged over the group elements, is the same as the average number of transpositions required to undo the permutations. We count fixed points as cycles of length one.

The key relation, of course, is the Cayley result, $T(\pi) = n - C(\pi)$, where $C(\pi)$ is the number of cycles induced by $\pi$, and $T(\pi)$ is the minimum number of transpositions required to reorder the permutation.

***Proposition B1***: Finite groups with the same cycle and transposition number of $\langle T \rangle = \langle C \rangle = \frac{n}{2}$, are balanced, and have polynomials with the same logarithmic derivatives at $x = 1$.

The proof is direct from the above since $\langle T_n(\pi) \rangle = n - \langle C_n(\pi) \rangle$, and by the definitions of the polynomials this is possible only if the averages of the number of transpositions and the number of cycles is *n*/2. From the formulas for these means in terms of the roots of the cycle polynomial for balanced groups we have

$$\sum_{k=1}^{n-1} \frac{1}{1 - r_k} = \sum_{k=1}^{n-1} \frac{-r_k}{1 - r_k} = \frac{n}{2}$$

(B1)

### Example of a balanced group

We now describe a simple example of a *balanced group*. Consider a permutation group acting on 4 objects. First, we note that the symmetric group $S_4$ is not a balanced group since $\langle T \rangle = 1.67$, and $\langle C \rangle = 2.08$. However, if any two of the elements consisting of a single transposition is deleted the resulting group becomes balanced, with $\langle T \rangle = 2$, and $\langle C \rangle = 2$. Note that since these elements are transpositions, and are their own inverses, the result of these deletions of group elements remains a group. Table 1 shows the relevant numbers for $S_4$, and a balanced version of this group is obtained by deleting two elements from the second type: one transposition elements.

|   | Cycle profile types | Size of conjugacy class | Cycle Number | Transp. Number |
|---|---|---|---|---|
| 1 | Four cycles of size one: four fixed points | 1 | 4 | 0 |
| 2 | One transposition (cycle of size two), & two fixed points | 6 | 3 | 1 |
| 3 | Two transpositions: two cycles of size two | 3 | 2 | 2 |
| 4 | One 3-cycle, one fixed point | 8 | 2 | 2 |
| 5 | One 4-cycle, no fixed points | 6 | 1 | 3 |

**Table 1.** The cycle and transposition profiles for the group $S_4$.

The polynomials for this balanced group can be read from the table modified by these deletions. Let us call the balanced versions of $S_4$, $SB_4$. Then the cycle and transposition polynomials are these:

$$P_{SB_4}(x) = x^4 + 4x^3 + 11x^2 + 6x \qquad (B2a)$$

$$Q_{SB_4}(x) = 6x^3 + 11x^2 + 4x + 1 \qquad (B2b)$$

A direct calculation shows that the roots for the $SB_4$ cycle polynomial are 0, and three complex roots. Larger balanced groups can similarly be constructed from the symmetric groups to much the same effect. It is clear, however, that for some values of $n$ such groups do not exist. Since the number of cycles and the number of reordering transpositions are reciprocal indications of order and disorder, the balanced groups are, in this sense at a central point between order and disorder.

The entropies induced by the balanced groups are interestingly simple. Expressing the integer entropy in terms of the average transposition numbers for a distribution of numbers $\{n_i\}$, such that $\sum_{i=1}^{m} n_i = N$, and a group $G_n$, we have the general expression

$$J_G = (N - \langle T \rangle_N) - \sum_{i=1}^{m} \frac{n_i}{N}(n_i - \langle T \rangle_{n_i})$$

(B3)

Now for a balanced group $\langle T_n \rangle = \langle C_n \rangle = \frac{n}{2}$, therefore as long as there are balanced groups for all numbers, we have the expression for the entropy

$$J_G(B) = \frac{N}{2} - \sum_{i=1}^{m} \frac{n_i^2}{2N}$$

(B4)

Now it is clear that if one of the *m* numbers is N and the rest zero the entropy is zero as expected, but if the distribution is uniform, $n_i = \frac{N}{m}$, then we have the simple form

$$J_G(B - uniform) = \frac{N}{2}\left(\frac{m-1}{m}\right)$$

(B5)

If the distribution is uniform over a given fraction, *f*, of the m numbers, while the rest are zero we have the following expression.

$$J_G(B - f) = \frac{N}{2}\left(\frac{m - f^{-1}}{m}\right)$$

(B6)

As we see from these examples the expression for the induced entropy can be rather simple. Further exploration of the use of balanced groups seems called for, and we will deal with this in another paper.